\documentstyle[preprint,aps]{revtex}

\newcommand\ignore[1]{}
\newcommand{\br}{{\bf r}}

\begin{document}

\draft
\title{Novel vortex lattice transition in $d$-wave superconductors}
\author{Jun'ichi Shiraishi${}^1$,
Mahito Kohmoto${}^1$ and Kazumi Maki${}^2$} 
\address{${}^1$Institute for Solid State Physics, 
University of Tokyo, Roppongi, Minato-ku, Tokyo 106, Japan}
\address{${}^2$Department of Physics and Astronomy,
University of Southern Calfornia Los Angeles,
Cal. 90089-0484, USA}

\maketitle
\begin{abstract}
We study the vortex state in a magnetic field parallel 
to the $c$ axis in the framework of the extended Ginzburg 
Landau equation. We find the vortex acquires a fourfold 
modulation proportional to $\cos(4\phi)$
where $\phi$ is the angle ${\bf r}$ makes with the $a$-axis. 
This term gives rise to an attractive interaction between 
two vortices when they 
are aligned parallel to $(1,1,0)$ or $(1,-1,0)$. 
We predict the first order vortex 
lattice transition at $B=H_{cr}\sim \kappa^{-1} H_{c2}(t)$ from 
triangular into the square lattice tilted 
by $45^\circ$ from the $a$ axis.
This gives the critical field $H_{cr}$ 
a few Tesla for YBCO and Bi2212 monocrystals 
at low temperatures ($T\leq 10 K$).

\end{abstract}
\pacs{ }
\narrowtext

The $d$-wave  superconductivity in the hole-doped high $T_c$ cuprate 
superconductors now appears to be universally established\cite{1,2}. 
Then in a magnetic 
field parallel to the $c$ axis we expect a square vortex lattice tilted
$45^\circ$ from the $a$ 
axis at least in the vicinity of $B=H_{c2}(T)$ and 
below $t( \equiv T/T_c)\leq 0.81$\cite{3,4}.
Indeed such vortex lattices though elongated in the $a$ direction 
have been 
seen by small angle neutron scattering (SANS)\cite{5}
and scanning tunnelling 
microscope (STM)\cite{6} both in twinned YBCO monocrystals 
at low temperature $(T \sim 4K) $
and in a small magnetic field ($B\sim $ a few Tesla). 
The object of this letter is to determine theoretically the
corresponding vortex lattice transition at a small magnetic field 
which is 
experimentally relevant.
Also in order this vortex lattice transition is to be observable
experimentally, it should take place in a magnetic field 
smaller than the one where vortex lattice melting 
is established\cite{7,8}.
Further we shall show that the transition is exthermal and 
of the first order. 
Earlier the related transition has been 
analyzed in terms of the generalized London equation, which may 
be obtained from the phenomenological Ginzburg Landau equation 
for $d$-wave superconductivity\cite{9} or from the rhombic term arising 
from the normal state properties\cite{10}. 
However these authors concentrated on the
modification of the magnetic interaction energy 
due to the quartic term in the kinetic energy. 
We believe, on the contrary, that the important modification lies in 
the vortex core structure\cite{11}.
Nevertheless these authors obtained the vortex 
lattice transition from a triangular lattice in small 
magnetic field to a  rhombic lattice in a  higher field. 
The result is similar to ours except for two crucial differences.
First we find that the critical field $H_{cr}$ where this 
transition takes place
behaves like 
$\kappa^{-1} H_{c 2}(t)$ where $\kappa$ is the Ginzburg Landau parameter 
and $H_{c2}(t)$ the upper critical field. 
This means $H_{cr}\sim  10^{-2} H_{c2}(t)$
in high $T_c$ cuprates. 
Second the transition is of the first  
order with a sharp negative divergence in the magnetization.

Within the weak-coupling model for $d$-wave superconductor\cite{12}
the extended Ginzburg-Landau equation is given by\cite{11}
\begin{eqnarray}
&&\left( 1 + 
 (\partial_x^2+\partial_y^2) 
+ \epsilon
\left[
5(\partial_x^2+\partial_y^2)^2+2(\partial_x^2-\partial_y^2)^2 \right]
\right) f(\br) =
|f(\br)|^2 f(\br). \label{GL0}
\end{eqnarray}
where we introduces the reduced unit $\br \rightarrow \br/\xi(T)$ ,
$\Delta(\br)=\Delta(T)f(\br)$,
\begin{eqnarray}
&&\xi(T)^2= { 7 \zeta(3)v^2\over 2 (4 \pi T)^2 (-\ln t)},\quad
\Delta(T)^2={  (4 \pi T)^2 (-\ln t)\over  21\zeta(3)}, 
\end{eqnarray}
and 
\begin{eqnarray}
&&\epsilon\equiv {31\zeta(5)(-\ln t)/ 196 \zeta(3)^2} \sim 0.114(-\ln t),
\end{eqnarray}
where $t=T/T_c$.
Here $\partial_x$ and $\partial_y$ are the gauge 
invariant differential operators.
Eq.(\ref{GL0}) has been obtained previously by 
Enomoto et al.\cite{13}, though we suppressed  
all the other terms of the order of $(-\ln t)$, which is 
not necessary for our purpose. 
Due to the quartic terms in Eq.(\ref{GL0})  the single 
vortex solution is now given by\cite{11}
\begin{eqnarray}
f(\br) &=& 
g(r) e^{i \phi}
+ \epsilon \left(
e^{4 i \phi} \alpha(r)+e^{-4 i \phi}\beta(r)+\gamma(r)\right) e^{i
  \phi}.
\end{eqnarray}
where $g(r)\sim \tanh r$ and $|f(\br)|$ and the phase
of $f(\br)$ are shown in Fig.1 and Fig.2. For
clarity we have chosen $\epsilon=1$. The ridges
in $|f(\br)|$ running along four diagonals are
clearly seen, which are crucial for the fourfold
core interaction energy. Also the phase distortion
propagates far away from the vortex center.
For $r\gg 1$, $\alpha(r)$ and $\beta(r)$ 
are obtained as \cite{11}
\begin{eqnarray}
\alpha(r)&=& {5 \over 2} r^{-2} + 
\left({5\over 4}-{55 \over 4} \log r \right) \; r^{-4}
+
\cdots, \label{alphainf}\\
\beta(r)&=& - {5 \over 2} r^{-2} +
\left({5 \over 4}+{55 \over 4} \log r  \right) r^{-4} 
+ \cdots,\label{betainf}
\end{eqnarray}
which tell us the explicit form of the asymptotics.
This result is consistent with Enomoto et al.\cite{13}.
In the presence of the fourfold distortion around single vortex, 
this term modifies drastically interaction between two vortices. 
Now let us consider the vortex lattice. Within the dimensionless
unit the free energy of the  vortex lattice is given by;
\begin{eqnarray}
\Omega&=&
\int d^2 r\; 
\left( -{1 \over 2} |\Delta|^4+{1 \over 8\pi} b^2
\right), 
\end{eqnarray}
where $b(\br)$ is the local induction. Making use of the usual 
approximation for $\Delta(\br)=\Delta \prod_i    f(\br-\br_i) $ 
where $\br_i$ is the position of 
the $i$-th vortex. We are here interested in the region 
$1 \ll d \ll\kappa$   
(or   $\xi(T) \ll d \ll \lambda(T) $
in the natural unit), where $d$ is the distance between 2 
neighboring vortices.
Then neglecting terms exponentially small we obtain
\begin{eqnarray}
\Omega&=&-{1 \over 2}\left(  A- a_1 \xi^2 n_\phi - 
\epsilon 10 a_1\xi^2 n_\phi
\sum_{l,m} {}' {\xi^4 \over r_{l,m}^4} \cos 4\theta_{l,m} \right)+
{2\pi\over \kappa^2} n_\phi \xi^2
\sum_{l,m}{}' K_0\left({r_{l,m}\over \lambda}\right),\label{omega}
\end{eqnarray}
where $A$ is the area, 
$a_1={8\pi \over 3}(\ln 2+{1 \over 8})\simeq 6.354$,
$n_\phi=B/\phi_0$ the vortex density per unit area and 
$\br_{l,m}=
2ld (\cos\theta,\sin\theta)+2md  (\cos\theta,-\sin\theta)$ or
$(2l+1)d (\cos\theta,\sin\theta)+(2m+1)d  (\cos\theta,-\sin\theta)$.
Here we limit ourselves to the vortex lattice with isoceles unit cell.
Also as to the magnetic interaction we consider here only the isotropic 
term $\kappa^{-2} K_0({r\over \lambda(T)}) $
since the anisotropic correction is much smaller than 
the vortex core interaction at 
least when $\kappa\gg 1$.
The last term in Eq.(\ref{omega}) is handled using 
the Poisson transform as in\cite{14}.
Finally we transform Eq.(\ref{omega}) as 
\begin{eqnarray}
\Omega&=&\Omega_0+{2 \pi \xi^2 H_{cr}\over \kappa^2 \phi_0} 
\overline{f}\left(B\over H_{cr}\right),
\end{eqnarray}
where  
\begin{eqnarray}
&&\Omega_0=
-{A\over 2} + {a_1 \xi^2 \over 2\phi_0} B+
{2\pi \xi^2 \over \kappa^2 \phi_0}B
\left[ 
{2\pi \lambda^2\over \phi_0} B+{1 \over 2}\log 
{\phi_0\over 2\pi \lambda^2 B}-{1 \over 2} (1-\gamma)
  \right],\\
&&\overline{f}\left(B\over H_{cr}\right)=
{B\over H_{cr}} 
f\left(\theta_{\rm min}\left({B\over H_{cr}}\right)\right),\\
&&f(\theta)=\left({B\over H^*(t)}\right)^2\sum_{l,m}{}' 
{\sin^2 2 \theta   \cos 4 \theta_{l,m} \over 
\left((l+m)^2 \sin^2 \theta+(l-m)^2 \cos^2 \theta\right)^2}\\
&&\quad + \sum_{l,m}{}' 
\left( E_1\left(\pi(l^2{\tan\theta }+
m^2{\cot \theta})\right)
+{(-1)^{l+m}+
 \exp \left(-\pi (l^2{\cot \theta}+m^2{\tan\theta})   \right)
  \over
\pi (l^2{\cot\theta}+m^2{\tan\theta})}
 \right),
\end{eqnarray}
and
\begin{eqnarray}
&&H^*(t)= \left({98 \zeta(3)^2 (2\pi)^3 \over 
155 a_1 \zeta(5) (-\ln t) }\right)^{1/2} {H_{c2}(t) \over \kappa}
\simeq 5.64667 (-\ln t)^{-1/2} {H_{c2}(t) \over \kappa}.
\end{eqnarray}
By minimizing $f(\theta)$ in $\theta$, we find the 
apex angle $\theta_{\rm min}$ shown in Fig.3.
$\theta_{\rm min}$ increases first gradually and 
jumps to $\pi/4$ at $B=H_{cr}$ where 
\begin{equation}
H_{cr}=0.524 (-\ln t)^{-1/2} \kappa^{-1} H_{c2}(t).
\end{equation}
Earlier a similar $\theta$-$B$ curve is obtained within 
generalized London 
equation \cite{9,10}. But here the arise of 
$\theta$ to $\pi/4$  is much steeper. 
Indeed contrary to \cite{9,10}, we predict that the transition is of 
the first order with a small dip in $-M$ \cite{11}.

Unfortunately no experimental data on 
$\theta_{\rm min}$ is available in the hole doped high $T_c$
cuprates to our knowledge.
However a very similar jump in 
$\theta_{\rm min}$ as function of $B$ as described here
has been seen by SANS from ErNi$_2$B$_2$C at $T=3.1$ \cite{15}.
Although the superconductivity of 
borocarbides is believed to be of $s$-wave,
both the existence of the rhombic vortex lattice
in a magnetic field $B$ parallel to the $c$ axis
seen by SANS \cite{15,16} in ErNi$_2$B$_2$C, YNi$_2$B$_2$C
and LuNi$_2$B$_2$C and STM from YNi$_2$B$_2$C \cite{17}
suggest strongly that the underlying supercondudtivity
is of $d$-wave. More recently
the fourfold anisotropy in the uppercritical field
in a magnetic field within the $a$-$b$ plane 
seen in LuNi$_2$B$_2$C \cite{18} is again consistent with $d$-wave
superconductivity \cite{19,20}.

To sum up we find that the vortex lattice is triangular as in a 
classical $s$-wave superconductor in a small 
magnetic field. As $B$ increases the vortex 
lattice transforms first gradually and then rapidly 
into a square lattice at $B= H_{cr}(t)$. 
Also the final jump is extremely rapid, which 
is seen from a sharp dip in the magnetization.

Although the theoretical expression for $H_{cr}(t)$ 
is obtained within the 
weak-coupling model and for $t$ not too small (say $t> 0.5$), from the 
temperature dependence of $C(t)$ in \cite{3}, we expect the 
result should be 
valid semiquantitatively even for $t\ll 1$, if we replace $-\ln t$ 
 by $1-t$. 
This allows us not only to estimate $H_{cr}(0)$ in the 
optimally doped YBCO and Bi2212 crystals but also to 
construct the phase diagram for the rhombic vortex lattice. 
As readily seen from Fig.4 the rhombic lattice occupies the
major part of the $T$-$B$ phase space of the vortex state unless
disturbed by the vortex lattice melting.

Though $H_{c2} (0)$ in YBCO or Bi2212 is not known 
for sure, we may assume $H_{c2}(0) \sim 
120$T  and $300$T for YBCO and Bi2212 respectively. 
Then making use of the fact    
$\kappa\simeq 100$ for these systems, we expect that 
the vortex lattice transition takes 
place at $B \sim 1$T and $3$T for YBCO and Bi2212 respectively.
The former is consistent with SANS and STM experiment in 
YBCO monocrystals mentioned 
in the beginning. For Bi2212, on the other hand, the melting 
transition appears to take 
place around $B\sim 300 $ {Gau\ss}  for optimally, doped Bi2212\cite{8}. 
If it is really the case, it will
be difficult to see the square lattice discussed here in Bi2212 
unfortunately. 
On the other
 hand if this transition exists, it will be not difficult to 
identify by SANS, STM and even 
by thermodynamic measurement.

\begin{center}
Acknowledgments
\end{center}

One of us (KM) thanks a very timely support of short term 
Japan Society of Promotion of 
Science fellowship which enables him to spend two weeks at 
ISSP at University of Tokyo. 
Also he is grateful to the hospitality and the support of 
Peter Wyder and MPI-CNRS at 
Grenoble where a part of this work is done. This work is in 
part of supported by 
NSF under grant member DMR9531720.

\newpage
\noindent
{\large \bf Figure Captions}

\begin{description}

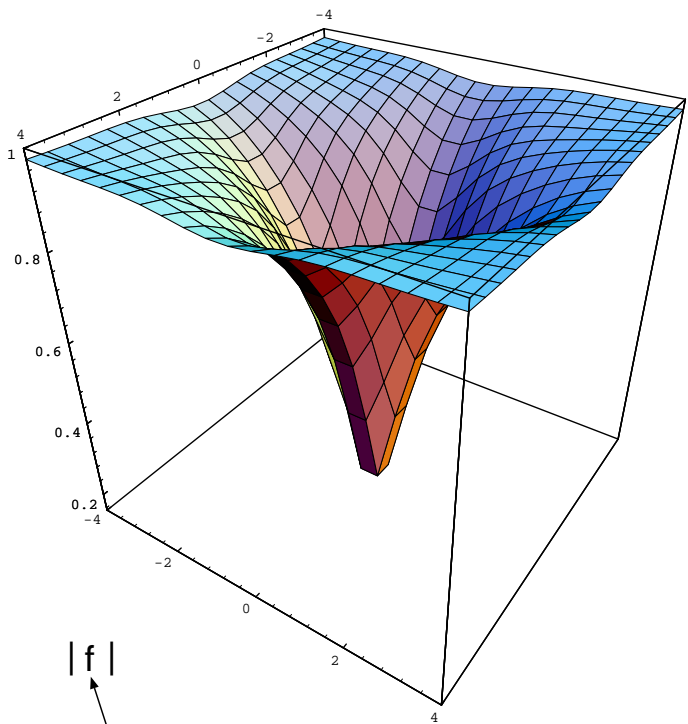
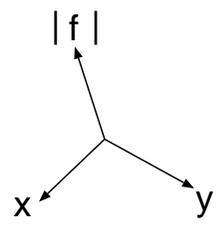
\item{Fig.1:}
$|f(\br)|$ in the $x$-$y$ plane. we used $\epsilon =1$
for the clarity of the figure, while $\epsilon \simeq 0.1$ in the real 
situation.

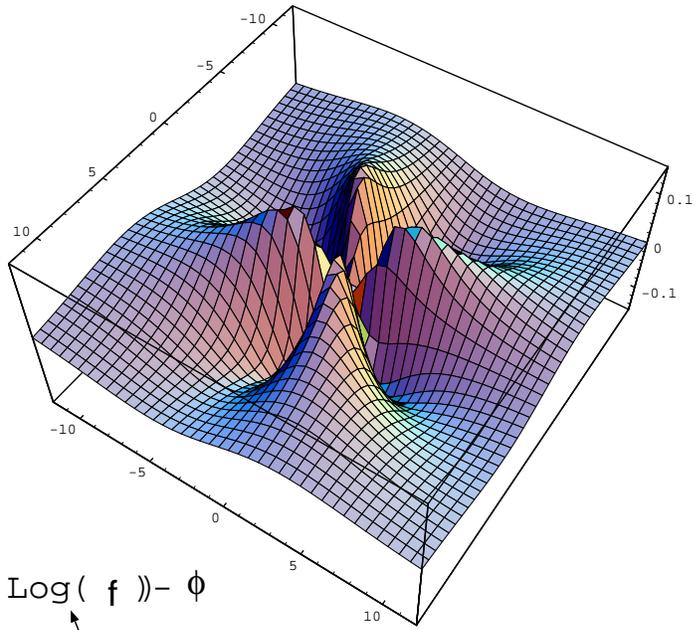
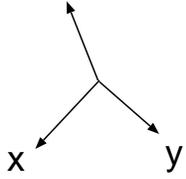
\item{Fig.2:}
$\Phi(\br)=\mbox{phase of $f(\br)-\phi$}$ is shown in the 
 $x$-$y$ plane for $\epsilon=1$.

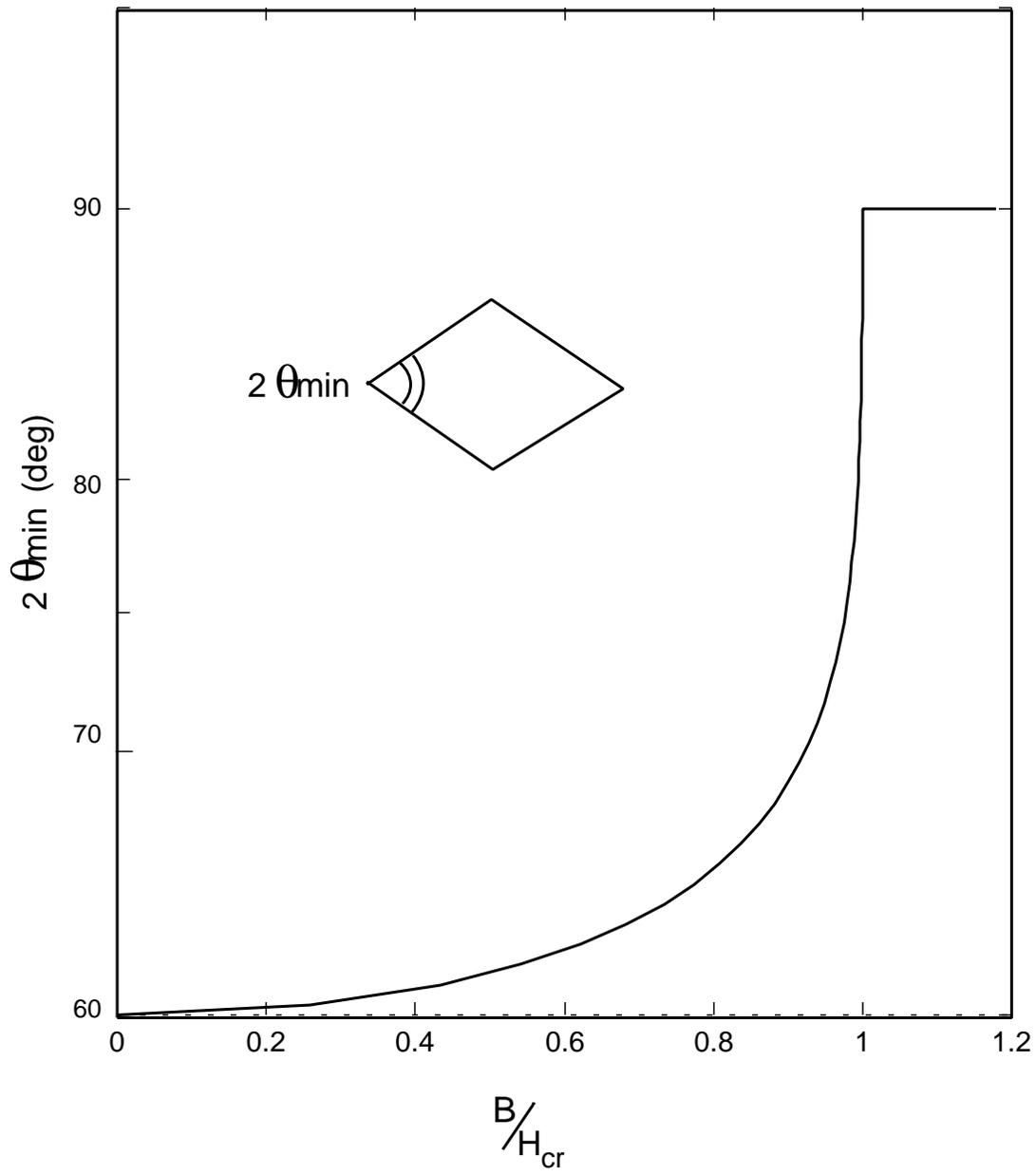
\item{Fig.3:}
Appex angle $2\theta_{\rm min}$  as a
function of $B/H_{cr}$ where $2\theta_{\rm min}=90^\circ$ and $120^\circ$ 
correspond to the square lattice and the triangular lattice with
hexagonal symmetry, respectively.

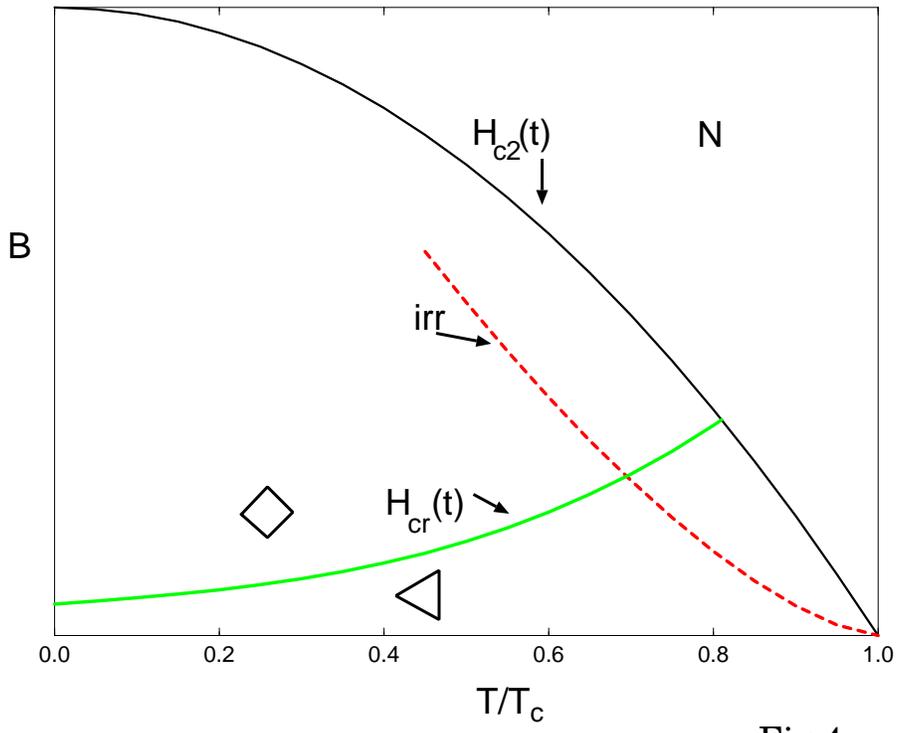
\item{Fig.4:}
The phase diagram with the rhombic vortex lattice is shown
schematically. In reality the rhombic lattice becomes stable around
$B=10^{-2} H_{c2}(t)$. Also ``irr'' means the irreducible
line where the vortex lattice melts into the vortex liquid by the
first order transition in the clean high $T_c$ cuprate
superconductor.

\end{description}

\end{document}